 \documentstyle[aps,eqsecnum]{revtex}
\begin{document}        
\vspace{0.3cm}        
\title{\bf\huge {Entropy computing via integration over fractal measures} }        
\vspace{0.5cm}        
\author{{\Large           
Wojciech S{\l}omczy{\'n}ski$^1$,        
Jaros{\l}aw Kwapie{\'n}$^2$,         
\\         
and Karol {\.Z}yczkowski$^{3,4}$}       
\\
\vspace{0.9cm}        
\begin{tabular}{lll}        
$^1$ & Instytut Matematyki,  Uniwersytet Jagiello\'nski, \\        
        &  ul. Reymonta 4, 30--059, Krak\'ow, Poland \\        
$^2$ & Instytut Fizyki J{\c a}drowej im.~H. Niewodnicza{\'n}skiego, \\        
        & ul. Radzikowskiego 152, 31--305, Krak\'ow, Poland \\        
$^3$ & Instytut Fizyki im.~M. Smoluchowskiego,        
Uniwersytet Jagiello\'nski, \\
        &  ul. Reymonta 4,  30--059, Krak\'ow, Poland \\
$^4$ & Centrum Fizyki Teoretycznej PAN  \\
     & Al. Lotnik{\'o}w 32/46, 02-668 Warszawa, Poland \\ 
\end{tabular}        
}        
\maketitle        
\vspace{0.4cm}        
\begin{center}        
{\small e-mail:        
$^1$slomczyn@im.uj.edu.pl \quad        
 $^2$kwapien@castor.if.uj.edu.pl \quad        
 $^3$karol@tatry.if.uj.edu.pl }        
\end{center}        
\vspace{0.4cm}        
\begin{center}        
   
\end{center}        
\vspace{0.4cm}        
\begin{center}        
{\sl Dedicated to the memory of Marcin Po{\'z}niak}        
\end{center}        
\vspace{0.4cm}        
        
\newpage        
        
\begin{abstract}        
{\large
We discuss the properties of invariant measures corresponding to        
iterated function systems (IFSs) with place-dependent probabilities        
and compute their R\'{e}nyi entropies, generalized dimensions, and        
multifractal spectra. It is shown that with certain dynamical        
systems one can associate the corresponding IFSs in such a way that        
their generalized entropies are equal. This provides a new method of        
computing entropy for some classical and quantum dynamical systems.        
Numerical techniques are based on integration over the fractal measures.        
  } 
\end{abstract}        
\newpage        
 
 {\bf
  In order to characterize quantitatively properties of a given
nonlinear system one often uses the notion of the dynamical entropy.
It describes the asymptotic changes of the system entropy in time.
Since analytical computing of this quantity is possible only for a limited
number of simple models,  it is important to develop efficient numerical
techniques for this purpose. In this article we propose a method
of computing the dynamical entropy  by averaging the
static Shannon entropy. The integration is performed
over a suitably chosen measure, which in the general case
displays  fractal properties.                                          
}
                                    
\section{Introduction} \label{s:introduction}        
        
{\sl Chaos} in a {\sl classical} dynamical system can be defined by        
the positiveness of the {\sl Kolmogorov-Sinai (KS) dynamical entropy}        
\cite{ER85}. This quantity, characterizing dynamical properties of        
a system, and defined via an asymptotic limit (time tending to        
infinity) is in general not easy to obtain analytically. On the        
other hand numerical computing of dynamical entropy from time        
series requires advanced techniques \cite{CP85,S88}. Also in the        
{\sl quantum} case estimating, so called, {\sl coherent states (CS)         
quantum entropy} \cite{SZ94,ZS95,KSZ97,SZ98} is not a simple task.         
In the present paper we propose a method of computing dynamical         
entropy of a system by finding an appropriate iterated function         
system with the same entropy.        
        
An {\sl iterated function system} (IFS) consists of a certain number        
$k$ of functions $F_i,~ i=1,\dots,k$, which act randomly with given        
probabilities $p_i, i=1,\dots,k$. An IFS may therefore be concerned        
as a combination of deterministic and stochastic dynamics. For        
sufficiently contracting functions one can prove (under some        
irreducibility conditions) that IFS generates a unique invariant        
measure (see Sect.~II). Generically \cite{LMy94,Sz98} this measure is        
localized on a fractal set. As it was described, e.g., in the elegant book of        
Barnsley \cite{B88} IFSs may be used to produce interesting fractal        
images, or to encode and transmit graphics via computer. For the majority        
of commonly analyzed and applied IFSs the probabilities $p_i$ are     
constant.        
For example, such IFSs have been used to construct multifractal energy        
spectra of certain quantum systems \cite{GM94} and to investigate the one-dimensional random-field Ising model \cite{BS88} or second order phase transitions \cite{R95}. On the other hand, with some classical and quantum dynamical systems one can associate in a natural way IFSs with place-dependent probabilities \cite{BDEG88,IG90,CGST93,LW93,GB95,KSZ97,S96}. In the present paper such IFSs will be called {\sl iterated function systems         
of the second kind}, on the analogy of position-dependent gauge         
transformations \cite{AL73}.        
        
We estimate the Kolmogorov-Sinai and R\'{e}nyi dynamical entropies of   
certain IFSs of the second kind using various numerical methods, which can   
be also applied in the general case. We use similar procedures to analyze   
the properties of the invariant measures of these IFSs and demonstrate their   
multifractal character. Eventually, we show that one can calculate the   
entropy of certain dynamical systems constructing IFSs with the same   
entropy. We give several examples that illustrate this new method of   
computing entropy.         
        
This paper is organized as follows. In the next section the        
definitions of IFSs of the first and second kind are recalled        
and their basic properties are considered. In Sect.~III we discuss        
briefly several methods of analytical and numerical computing of        
dynamical entropy of an IFS. In Sect.~IV we study generalized        
dimensions of measures which are invariant under the action of a        
one-dimensional IFS. Sect.~V presents a detailed analysis of a family         
of IFSs of the second kind and their invariant measures. Certain         
integrals over these measures are calculated. Moreover we compute         
the R\'{e}nyi entropies, generalized dimensions and multifractal spectra         
of these measures. In the subsequent section we investigate the         
connection between one-dimensional dynamical systems and IFSs of         
the second kind. In particular, IFSs associated to asymmetric tent map,         
logistic map, and "hut map" are analyzed and their entropies are         
calculated. We also show how one can apply this method to compute         
CS-quantum entropy. Concluding remarks are contained in        
the last section.        
        
In this paper we present only the results and numerical calculations. For the        
proofs we refer the reader to a forthcoming publication.         
        
\section{ Iterated function systems and their invariant measures}        
        
An iterated function system (IFS) is specified by $k$ functions        
transforming a metric space into itself and $k$ place-dependent        
probabilities which characterize the likelihood of choosing a        
particular map at each step of the evolution of the system. Under        
certain contractivity and irreducibility conditions one can prove the        
existence of a unique attractive invariant measure for an IFS, as well        
as ergodic and central limit theorems. Miscellaneous results of this        
type have been established since late thirties (some of them have        
been proved independently by several authors) - see for instance        
\cite{K81,E87,BE88,BDEG88,IG90,AC92,G92,G94,LY94,JL95,L95,LR95,FM98,SS98,O98}         
and references therein.        
        
In the present paper we study IFSs~ ${\cal F} = \{F_i,        
p_i : i=1,\dots,k\}$ that fulfil the following (rather strong)        
assumptions which guarantee veracity of the above mentioned        
theorems:        
        
\vspace{1 mm} {\bf General Assumption:}        
        
\vspace{1.5mm} {\bf (1)} $X$ is a compact metric space;        
        
\vspace{1.5mm} {\bf (2)} $F_i : X \to X,~~i=1,\dots,k$ are        
Lipschitz  functions with the Lipschitz constants $L_i < 1$;        
        
\vspace{1.5mm} {\bf (3)} $p_i : X \to [0,1],~~i=1,\dots,k$ are        
H\"{o}lder continuous functions fulfilling        
$\sum_{i=1}^{k} p_i(x) = 1$ for each $x\in X$;        
        
\vspace{1.5mm} {\bf (4)} $p_i(x) > 0$ for every $x\in X$ and $i =        
1,\dots,k$.        
        
\vspace{1.5mm} Such IFSs are often called {\sl hyperbolic}.         
Unless otherwise stated we assume that all IFSs under consideration         
are hyperbolic.        
        
Let us recall briefly several basic facts on IFSs.        
        
The IFS ${\cal F} = \{F_i,  p_i : i=1,\dots,k\}$ generates the        
following {\sl Markov operator} $V$ acting on $M(X)$ (the space        
of all probability measures on X):        
        
\begin{equation}        
(V\nu)(B) = \sum_{i=1}^{k} \int_{F_i^{-1}(B)} p_i(\lambda)        
d\nu(\lambda),        
\label{IFS1}        
\end{equation}        
where $\nu \in M(X)$ and $B$ is a measurable subset of $X$. This        
operator describes the {\sl evolution of probability measures} under        
the action of ${\cal {F}}$. The related {\sl Markov process} can be        
defined in the following way. As a probability space we take        
the {\sl code space}         
$\Omega ={\{1,\dots,k\}}^{{I\!\!N}}$        
and we put  $P_x$ for the probability measure on $\Omega$ given by        
        
\begin{eqnarray}        
P_x(i_1,\dots,i_n) &:=&        
P_x(\{\omega \in \Omega : \omega(j) = i_j, j=1,\dots,n\}) \\ \nonumber        
&:=&        
p_{i_1}(x)p_{i_2}(F_{i_1}(x)) \cdots p_{i_n}( F_{i_{n-1}}(        
F_{i_{n-2}}(\dots(F_{i_1}(x))))),        
\label{IFS2}        
\end{eqnarray}        
where $x \in X, i_j = 1,\dots,k, j = 1,\dots,n; n \in {I\!\!N}$.        
Then the formulae        
        
\begin{equation}        
Z_n^x(\omega) = F_{\omega(n)}( F_{\omega(n-1)}        
(\dots(F_{\omega(1)}(x)))), ~ ~ ~    Z_0^x(\omega) = x        
\label{IFS3}        
\end{equation}        
for $x \in X, \omega \in \Omega, n \in {I\!\!N}$        
define the requested Markov stochastic process on $(\Omega,        
\{P_x\}_{x \in X})$.

One can show that for an IFS which fulfils our assumption there        
exists a unique {\sl invariant probability measure} $\mu$ satisfying        
the equation $V\mu = \mu$ (the proof of this claim can be found in \cite{BDEG88}). This measure is {\sl attractive}, i.e.,        
$V^n{\nu}$ converges weakly to $\mu$ for every $\nu \in M(X)$        
if $n \to \infty$ or, in other words, $\int_X u~d V^n{\nu}$ tends         
to $\int_X u~d{\mu}$ for every continuous $u : X \to {I\!\!R}$.         
Thus, in order to obtain the exact value of $\int_X u~d{\mu}$, it         
is sufficient to find the limit of the sequence $\int_X u~d V^n{\nu}$         
for an arbitrary initial measure $\nu$. For instance, taking $\nu$ equal         
to a Dirac delta measure $\delta_x$ for some $x \in X$ we obtain the        
integral of $u$ over the invariant measure $\mu$ as the limit of the        
sequence        
        
\begin{equation}        
U_n: = \sum_{i_1,\dots,i_n=1}^k P_x(i_1,\dots,i_n)        
u(x_{i_1,\dots,i_n} ),        
\label{IFS4}        
\end{equation}        
where $x_{i_1,\dots,i_n} := F_{i_n}( F_{i_{n-1}}(        
\dots(F_{i_1}(x))))$. After Barnsley \cite{B88} (see also        
\cite{E96}) we call this method of computing integrals over the        
invariant measure {\sl deterministic algorithm}. To find the integral        
numerically we can also employ the ergodic theorem for IFSs        
\cite{K81,E87,IG90,G94,FM98,SS98}. Any initial point $x \in X$ iterated         
by the IFS generates a random sequence         
$(z_0~=~x, z_1,\dots,z_n ,\dots)$, where $z_i := Z_i^x(\omega)$.         
Then        
        
\begin{equation}        
\int_X u(x) d\mu(x) =  \lim_{n\to\infty} {1\over n}        
\sum_{i=0}^{n-1} u(z_i)        
\label{IFS5}        
\end{equation}        
with probability one, i.e., except of a set of measure        
$P_x$ zero. Moreover, if $u$ fulfils the Lipschitz condition we can        
evaluate the rate of convergence in the ergodic theorem utilizing the        
central limit theorem for IFS \cite{IG90,LR95}. This leads to        
a probabilistic (Monte-Carlo) numerical method of computing        
integrals over the invariant measure which was called {\sl random 
iterated algorithm} by Barnsley \cite{B88}. We have successfully        
applied both techniques to compute numerically various integrals,        
including those necessary to estimate the dynamical entropy of IFS        
(see Sect.~III).        
        
Finally, let us look at the evolution of densities under the action of        
IFS. If $m$ is a finite measure on $\Omega$ and $F_i~~        
(i=1,\dots,k)$ are {\sl nonsingular} (i.e., $m(A)=0$ implies        
$m\Bigl(F_i^{-1}(A)\Bigr) = 0$ for each measurable $A \subset        
X$), then the IFS $\cal{F}$ generates a  Markov operator on the        
space of densities (with respect to $m$) on $X$, also called the {\sl        
Frobenius-Perron operator} \cite{LM94}. It is so if, for instance,        
$X$ is an interval in ${I\!\!R}$ and  $\{F_i:i=1,\dots,k\}$ are        
diffeomorphisms. It follows from (\ref{IFS1}) that the        
Frobenius-Perron operator $M$ associated with $\cal{F}$ is given        
in this case by the formula        
        
\begin{equation}        
M[\gamma](x) = \sum_{i} p_i\Bigl(F_i^{-        
1}(x)\Bigr)\gamma\Bigl(F_i^{-1}(x)\Bigr)        
\left| \frac{d F_i^{-1}(x)}{dx} \right|,        
\label{IFS6}        
\end{equation}        
where the sum goes over all $1 \leq i \leq k$ such that $x \in        
F_{i}(X)$, for $\gamma$ a density and $x \in X$.        
        
 If probabilities $p_i$ are constant then we will say that an IFS is {\sl        
of the first kind}. IFSs with place-dependent probabilities will be        
called IFSs {\sl of the second kind} (they also appear in the literature under     
the name of {\sl learning systems}).

\section{Entropy of IFS}        
        
Let $\mu$ be the attractive invariant measure for the IFS ${\cal F} =        
\{F_i, p_i : i=1,\dots,k\}$. We define the probability measure        
$P_{\mu}$ on the code space $\Omega = {\{1,\dots,k\}}^{{I\!\!N}}$        
by        
        
\begin{equation}        
P_{\mu}(i_1,\dots,i_n) :=        
P_{\mu}(\{\omega \in \Omega : \omega(j) = i_j, j=1,\dots,n\}) :=        
\int_X P_x(i_1,\dots,i_n) d{\mu}(x)        
\label{ren1}        
\end{equation}        
for $i_j = 1,\dots,k, j = 1,\dots,n; n \in {I\!\!N}$.        
        
It is easy to show that this measure is invariant with respect to the        
shift on $\Omega$.        
        
Now we can define the {\sl partial entropies} as        
        
\begin{equation}        
H(n):= - \sum_{i_1,\dots,i_n=1}^{k}        
 P_{\mu}(i_1,\dots,i_n ) \ln P_{\mu}(i_1,\dots,i_n ),        
\label{hn}        
\end{equation}        
and the {\sl relative entropies} by        
        
\begin{equation}        
G(1):= H(1); ~ ~  G(n)=H(n)-H(n-1),~ {\rm for}~ n>1.        
\label{gn}        
\end{equation}        
The dynamical entropy of Kolmogorov and Sinai can be extracted        
from both sequences, i.e.,        
$K_1=\lim_{n\to \infty} G(n) = \lim_{n\to \infty} H(n)/n$.        
The usage of relative entropies is often advantageous, since the        
convergence of $H(n)/n$ is slow (usually as $1/n$), while in many        
cases the sequence $G(n)$ converges to the KS-entropy        
exponentially fast \cite{CP85,ZS95b}. Note that the entropy of a        
stochastic system like an IFS can be defined in several different        
ways \cite {K86}. Here we are interested in the dynamics induced by        
an IFS in the $k$-symbols code space, which leads to the        
entropy finite and bounded by ln$k$.        
        
The concept of dynamical KS-entropy is based on the notion of the        
Boltzmann-Shannon entropy function which can be multifariously        
generalized \cite{K95}. In this paper we discuss the two versions of        
R\'{e}nyi entropy defined in \cite{KSZ97} for any real $q \ne 1$.        
The first one, often used in the literature, corresponds to the limit of        
{\sl partial entropies}:        
        
\begin{equation}        
\tilde{K}_q := \limsup_{n \to \infty} {\frac {1}{n}}        
{\frac{1 }{1 - q}} \ln \Bigl[~\sum_{i_1,\dots,i_n=1}^k        
[P_{\mu}(i_1,\dots,i_n)]^q \Bigr].        
\label{ren2}        
\end{equation}        
The other one, based on the notion of R\'{e}nyi conditional entropy        
\cite{R61}, is defined via {\sl relative entropies}:        
        
\begin{equation}        
K_q := \limsup_{n \to \infty}        
{\frac{1 }{1 - q}} \ln \Bigl[\sum_{i_1,\dots,i_n=1}^{k}        
 P_{\mu}(i_1,\dots,i_n )\Bigl(P_{\mu}(i_1,\dots,i_n ) /        
P_{\mu}(i_1,\dots,i_{n-1})\Bigr)^{q-1}\Bigr].        
\label{ren3}        
\end{equation}        
For $q=0$ both quantities are equal to the topological entropy        
$K_0=\ln k$ and the KS-entropy is obtained  for both quantities        
in the limit        
$q \to 1$. On the other hand, in general, the two versions of        
R\'{e}nyi entropies are different (see Sects.~III~and~V). The        
computation of entropy $K_q$ is more straightforward than        
$\tilde{K}_q$ and an analytical treatment is possible in some cases,        
on the other hand, its relation to the thermodynamical formalism        
seems to be less clear.         
        
Both definitions give us some form of the R\'{e}nyi dynamical entropy for        
the dynamics generated by an IFS in the $k$-symbols code space and for        
the specific partition of this space into $k$ rectangles labeled by the first     
symbol. Note, however, that if one defines the (partition independent)     
R\'{e}nyi dynamical entropy taking simply the supremum over all finite     
partitions (as for the KS-entropy), this leads to trivial dependence:     
$K_q=\infty,\;q<1;\;\;K_q=K_1,\;q \geq 1$ \cite{T98}. Consequently, from     
now on, we shall discuss only the R\'{e}nyi entropy for the above mentioned     
$k$-elements partition.      
        
For an IFS of the first kind (with constant probabilities $p_i$)        
both R\'{e}nyi entropies are equal and can be written        
down explicitly \cite{BPTV88,GOTV93}        
        
\begin{equation}        
K_q = \tilde{K}_q = {1 \over 1-q} \ln        
(p_1^q+p_2^q+\cdots+p_k^q),        
\label{ren4}        
\end{equation}        
for $q \neq 1$. The KS-entropy is obtained by        
calculating the limit $\lim_{q\to 1} K_q$, which gives        
$K_1 = \tilde{K}_1 = -\sum_{i=1}^k p_i \ln{p_i}$. Observe that        
for an IFS of the first kind the value of the entropy does not depend        
on the character of functions $F_i$.        
        
For an IFS of the second kind one cannot directly apply formula        
(\ref{ren4}), since the probabilities are place-dependent.        
A natural generalization for this case is possible \cite{KSZ97,S98},        
viz., one has to average the R\'{e}nyi entropy performing an integral        
over the invariant measure $\mu$        
        
\begin{equation}        
K_q = {1 \over 1-q } \ln \int_X \sum_{i=1}^{k}        
(p_i(x))^q  d{\mu}(x).        
\label{ren5}        
\end{equation}        
In the limit $q \to 1$, corresponding to KS-entropy, this formula        
gives        
        
\begin{equation}        
K_1 =  - \int_X \sum_{i=1}^{k}        
p_i(x) \ln[p_i(x)]  d{\mu}(x).        
\label{ren6}        
\end{equation}        
Moreover, one can show that the relative entropies converge to the        
limiting value $K_q$ exponentially \cite{S98}. Now, to compute the        
entropy, it suffices to apply one of the two methods of calculating the        
integral over the invariant measure of an IFS presented in Sect.~II.        
        
To estimate entropy $\tilde{K}_q$ one may consider the modified        
IFS ${{\cal {F}}_q} = \{F_i, \tilde{p}_i(q) : i=1,\dots,k\}$ with the        
probabilities $\tilde{p}_i(q)$ proportional to ${p_i}^q$, that is,          
given by the formula        
        
\begin{equation}        
\tilde{p}_i(q)(x) = (p_i(x))^q /  \sum_{j=1}^{k}(p_j(x))^q.        
\label{ren7}        
\end{equation}        
for $x \in X, i = 1,\dots,k$, and $q \in {I\!\!R}$.        
        
Note that a similar method was used for one-dimensional IFSs of the first kind in \cite{BS88}. It is easy to prove that the IFS ${\cal {F}}_q$ satisfies our general assumption, and hence, a unique invariant probability measure $\mu^q$ exists. Then one can derive \cite{S98} the following inequality        
        
\begin{equation}        
(1-q) \tilde{K}_q  \geq \int_X \ln \sum_{i=1}^{k}        
(p_i(x))^q  d{\mu}^q (x),        
\label{ren8}        
\end{equation}        
which provides a lower bound for entropy $\tilde{K}_q $ for        
$q < 1$, and an upper bound for $q>1$. In examples we analyze         
(see Sects.~V.C~and~VI.B) this bound is actually very close to the         
exact value of the entropy $\tilde{K}_q$ calculated numerically.         
Furthermore, the integral on the right-hand side of  (\ref{ren8}) can         
be relatively easily computed (see Sect.~II), whereas the convergence         
in (\ref{ren2}) seems to be rather slow, namely as $n^{-1}$.        
        
\section{Dimensions of invariant measure for IFS}        
        
In this section we assume that a one-dimensional ($X \subset {I\!\!R}$)        
IFS ${\cal F} = \{F_i, p_i : i=1,\dots,k\}$ is given, where $ F_i $ are         
diffeomorphisms fulfilling the general assumption        
from Sect.~II and the following {\it separation condition}:        
$intF_i(X) \cap intF_j(X) = \emptyset$ for $i \neq j, i,j = 1, \dots, k$,         
where $intF_i(X)$ denotes the interior of the set $F_i(X)$.        
Our aim is to calculate the {\it generalized dimensions} $D_q$ of the        
invariant measure for ${\cal F}$. These quantities were introduced  and        
analyzed by Grassberger, Hentschel, and Procaccia \cite{HP83,GP83a}         
(for more information see \cite{O94,O95,Fa97}), and $D_0$ is just         
the Hausdorff dimension of the invariant measure. The correlation         
dimension $D_2$ for certain  IFSs has been recently studied  by        
Chin, Hunt and Yorke \cite{CHY97}.        
        
Let us consider the following {\it pressure function}        
        
\begin{equation}        
P(q,\tau )(x) =: \limsup_{n \to \infty} {\frac {1}{n}} \ln        
\Bigl[~\sum\limits_{i_1,...,i_n=1}^kP_x(i_1,...,i_n)^q\left| \left( F_{i_n}        
\circ...\circ F_{i_1}\right) ^{\prime }\left( x\right) \right| ^{-\tau }\Bigr]        
\label{dim1}        
\end{equation}        
for $q\geq 0$ and $\tau \in {I\!\!R}$.        
        
In the sequel we assume that the limit in (\ref{dim1}) does not         
depend on $x$ and the generalized dimensions $D_q$ are given by the         
formula        
\begin{equation}        
D_q={\tau(q) \over q-1}~,        
\label{dim2}        
\end{equation}        
where $\tau(q)=\tau$ is the only solution of the equation $P(q,\tau)=0$.        
For the IFS of the first kind this assumption was heuristically verified by        
Halsey et al. \cite{HJKPS86} (see also \cite{O95}). Moreover Bohr         
and Rand \cite{BR87,R89} showed that it holds for the IFS generated         
by expanding maps on the interval ("cookie-cutters").        
        
To estimate the generalized dimension $D_q$ we can use the technique        
already introduced in the preceding section. Namely, we consider the        
modified IFS ${\cal {F}}_{q,\tau} = \{F_i, \tilde{p}_i(q,\tau) :        
i=1,\dots,k\}$ with the probabilities $\tilde{p}_i(q,\tau)$ given by        
        
\begin{equation}        
\tilde{p}_i(q,\tau)(x) = (p_i(x))^q\left|F_i ^{\prime }(x)\right|^{-\tau}/        
\sum_{j=1}^{k}(p_j(x))^q\left|F_j ^{\prime }(x)\right|^{-\tau}.        
\label{dim3}        
\end{equation}        
for $x \in X, i = 1,\dots,k$, $q>0$, and $\tau \in {I\!\!R}$.        
        
Again it is easy to prove that the IFS ${\cal {F}}_{q,\tau}$ fulfils our        
general assumption, and hence admits a unique invariant probability        
measure $\mu^{q,\tau}$. Then one can show \cite{S98} that the        
following inequality holds        
        
\begin{equation}        
(q-1) D_q  \leq (q-1){\overline{D}_ q},        
\label{dim4}        
\end{equation}        
where $(q-1)\overline{D}_ q = {\overline{\tau}}$ is the solution of         
the equation        
        
\begin{equation}        
\label{dim5}        
\int_X \ln \sum_{i=1}^{k}        
(p_i(x))^q\left|F_i ^{\prime }(x)\right|^{-\overline{\tau}}d{\mu}^        
{q,\overline{\tau}}(x) = 0,        
\end{equation}        
for $q >0$.        
        
This provides a lower bound for the generalized dimension $D_q$ for        
$q < 1$, and an upper bound for $q>1$. In all the cases we study in         
Sects.~V.C~and~VI.B  this bound (which can be relatively easily         
computed) is actually very close to the value of the dimension         
$D_q$ calculated numerically. In order to calculate the generalized         
dimensions $D_q$ we use the "box-counting" algorithm,         
which in this case yields better results than the algorithm of         
Grassberger and Procaccia \cite{GP83a,S88} applied to the time series         
extracted from the IFS. Note that if  $\left|F_i ^{\prime }(x)\right| \equiv         
L>0$ for all $i =1,\dots,k$, then the generalized entropies and dimensions         
are related by a simple formula $\tilde{K}_q =- D_q \ln{L}$ (a relation between both quantities for IFSs of the first kind was examined in \cite{BS88}).        
        
Scaling properties of the invariant measure could be described with        
the aid of its multifractal spectrum        
$f(\alpha) = \inf_q \{\alpha q +(1-q)D_q\}$        
(for more information on multifractal spectrum see         
\cite{S88,O94,O95,BPS97}).

\section{Multifractals generated by IFS of the second kind}        
        
\subsection {Cantor measures}        
        
Let us consider a family of IFSs        
$\{X=[0,1],k=2;\ F_1\left( x\right) = x/3,\ F_2\left(x\right)=(x+2)/3;        
\  p_1(x) = (1-a)+(2a-1)x,\  p_2(x) = a + (1-2a)x ~~\hbox {for}~~ x        
\in X \}$, where $a \in [0,1]$.        
It is easy to see that these IFSs fulfil our general assumption for $a        
\in (0,1)$, which guarantees the existence of a unique invariant        
measures $\mu_a$ and veracity of the other results mentioned in        
Sects.~II,~III,~and IV. For $a=1$ one can prove the existence of a         
unique attractive invariant measure as well as the ergodic theorem         
(but not the central limit theorem) using more refined results which         
may be found in \cite{IG90} or \cite{G94}. On the other hand,         
for $a=0$, the IFS attracts every measure into a linear combination         
of two Dirac deltas localized at points $0$ and $1$. Hence there         
exists a whole family of invariant measures         
$\{r\delta_0 + (1-r)\delta_1\ : r \in [0,1]\}$ in this case.        
        
An IFS of the first kind is obtained for $a=1/2$, since the        
probabilities $p_1(x) = p_2(x) \equiv 1/2$ do not depend on $x$.        
The invariant measure $\mu_{1/2}$ is spread uniformly over the        
Cantor set. The generalized fractal dimension is constant $D_q =        
D_0 = \ln2/\ln3$, which implies a singular multifractal spectrum        
concentrated at  $\alpha _1=\ln2/\ln 3$ with $f(\alpha_1)=\alpha_1$.        
The generalized R{\'e}nyi entropy for this IFS can        
be directly obtained from (\ref{ren4}). It gives $K_q = \ln2$ for all        
$q \in {I\!\!R}$. The Cantor measure $\mu_{1/2}$ can thus be called        
both {\sl uniform} (constant generalized dimension) and {\sl        
balanced} (constant R\'{e}nyi entropy) \cite{BPTV88}.        
        
In the case $a=1$ the probabilities are place-dependent        
($p_1(x)=x; p_2(x)=1-x$) and define an IFS of the second kind. In        
order to understand the nature of the measure $\mu_1$, let us        
consider the iterations $\gamma_n = M_1(\gamma_{n-1})$ of the        
initially uniform density $\gamma_0$ with respect to the        
Frobenius-Perron operator $M_1$ given by (\ref{IFS6}).        
        
We simplify the notation by introducing the 'box' functions $x \to        
\Theta_{a,b}(x):=\Theta (x-a)\Theta (b-x)$, with the Heaviside function $\Theta$ given by $\Theta(y)=0$, for $y<0$, and $\Theta(y)=1$, for $y\geq 0$. The uniform density in $X$ can thus be written as $\gamma_0=\Theta_{0,1}$.        
        
Formula (\ref{IFS6}) allows us to obtain for instance the first two        
iterations of $\gamma_0$        
        
\begin{equation}        
\gamma_1 (x)=9x\Theta_{0,{1 \over 3}}(x)        
+9(1-x)\Theta_{{2\over 3},1}(x),        
\label{nu1}        
\end{equation}        
and        
        
\begin{equation}        
\gamma_2\smallskip (x)=3^5x^2\Theta_{0,{1\over 9}}(x)        
+3^4(x-3x^2)\Theta_{{2\over 9},{3\over 9}}(x)        
+3^4(3x-2)(1-x)\Theta_{{6\over 9},{7\over 9}}(x)        
+3^5(1-x)^2\Theta_{{8\over9},1}(x)        
\label{nu2}        
\end{equation}        
for $x \in [0,1]$.        
        
Similarities and differences between densities approximating the        
standard Cantor measure $\mu_{1/2}$ and the measure        
$\mu_1$ are displayed in Fig.~1. Due to constant probabilities, in        
the first case the measure $\mu_{1/2}$ covers uniformly the Cantor        
set (Fig. 1a, 1b and 1c). On the other hand, for the IFS of the second        
kind, the place-dependent probabilities induce a highly non-uniform        
distribution of the measure $\mu_1$ (Fig. 1d, 1e and 1f). For         
example, in each connected component of the support of the density         
$\gamma_n$ it can be expressed as a polynomial in $x$ of $n$-th         
degree. Note that, if $\gamma_n$ achieves its maximum at $x_n$,         
then $\gamma_{n+1}(x_n)=0$. One may expect, therefore, that         
the measure $\mu_1$ is multifractal.        
        
\subsection {Integration over fractal measures}        
        
Let us now calculate the integrals of certain        
functions $u$ over the invariant measures $\mu_{1/2}$ and        
$\mu_1$. Let us find, for example, the mean $(u_A(x)=x)$ and the        
mean square $(u_B(x)=x^2)$ for these measures. Iterating the        
uniform density $\gamma _0$ by the Frobenius-Perron operator        
$M_{1/2}$ we get the sequences of integrals        
${{U_A}_n}^{(1/2)} = \int_X u_{A(x)} \gamma _n(x)dx = 1/2$         
(independently of $n$) and ${{U_B}_n}^{(1/2)} = \int_X u_B(x)         
\gamma _n(x)dx = 3(1-3^{-2n-2})/8$. Consequently, two integrals         
in question read $\int_X x d\mu_{1/2}(x)=1/2$ and $\int_X x^2        
d\mu_{1/2}(x) = 3/8$. Computing integrals over the measure        
$\mu_1$ it is advantageous to start with an initially singular        
measure. To demonstrate the convergence rate explicitly we take a        
one parameter family of measures consisting of a combination of        
two delta peaks localized in both ends of the unit interval: $\kappa_r        
= r\delta _0+ (1-r)\delta_1$, where $r\in [0,1]$. Iterating this        
measure with respect to the Markov operator  $V_1$ given by        
(\ref{IFS1}) we compute the $r$-dependent integrals of both        
functions        
${{U_A}_n}^{(1)} = {1\over 2}[1 +(-{1\over 3})^n ] -r (-{1\over 3})^n        
\rightarrow \frac 12$        
and        
${{U_B}_n}^{(1)} = {1\over 3}[1 +2(-{1\over 3})^n ] -r (-{1\over        
3})^n        
\rightarrow \frac 13$.        
Both sequences tend to their limits independently of the parameter        
$r$, which contributions into the integral decay with $n$ as        
$3^{-n}$. Since both measures $\mu_{1/2}$ and $\mu_1$ are        
symmetric with respect to $x=1/2$, the first moments        
$\langle x\rangle$ are equal, however, already the second moments        
reveal the difference.        
        
In a similar  way an integral of a function over the Cantor set        
may be expressed as a limit of the sum of $2^n$ terms (multiplied by the        
appropriate weights), which probe the function on the ends of the        
intervals composing the Cantor set. In some cases this result can        
be put into a form of an infinite product. For example the        
characteristic function of the uniform Cantor measure        
$\mu_{1/2}$ is given by \cite{BBW97} (see also \cite{H96})        
        
\begin{equation}        
\int_0^1 e^{itx} d\mu_{1/2}(x)  = e^{it/2} \prod_{n=1}^{\infty}        
\cos ({t \over 3^n }).        
\label{prod}        
\end{equation}        
Due to fast convergence this form is particularly useful for numerical        
evaluation. In general, computing the integrals over multifractal        
measures generated by IFSs of the second kind one has to rely on         
numerical methods described in Sect.~II. For IFSs with small number         
of functions the deterministic algorithm based on (\ref{IFS4})         
provides more precise results than the random iterated algorithm         
(\ref{IFS5}). The latter seems to be more efficient for IFSs         
consisting of many functions.        
        
\subsection {Computing of entropies and dimensions via integration        
over the fractal measures}        
        
The entropy of the IFSs can be expressed as an integral of the        
R\'{e}nyi (or Boltzmann-Shannon) entropy function over the        
invariant measure $\mu_a$ (see (\ref{ren5}) and (\ref{ren6})) for        
$K_q$, or estimated by the respective integral over the measure        
$\mu^q_a$ (see (\ref{ren8})) for $\tilde{K}_q$. Note that, comparing         
the latter case with the former, the natural logarithm interchanges with         
the integral over $X$.        
        
Numerically computed R\'{e}nyi entropies $K_q$ and $\tilde{K}_q$        
of the measure $\mu_1$ are displayed on Fig.~2a (however,        
formula (\ref{ren8}) is valid only for $q > 0$ in this case). As expected,         
both entropies depend substantially on the R\'{e}nyi parameter $q$,        
which means that the invariant measure $\mu_1$ is not balanced.        
Note that the inflection point of the curve $K_q$ is situated not at        
$q=0$ but at some negative $q_c$. Making use of the integrals        
${U_A}_n(1)$ and ${U_B}_n(1)$ we obtain analytical results        
$K_2 = (\ln 3)/2$ and         
$K_3 = (\ln 2)/2$ directly from (\ref{ren5}).        
R\'{e}nyi entropies allow one to compute the        
scaling spectra via the Legendre transform:        
$g(\alpha) = \inf_q \{\alpha q + (1-q)K_q\}$ and $\tilde{g}(\alpha)        
= \inf_q \{\alpha q + (1-q)\tilde{K}_q\}$ (see Fig.~2b). The common        
maximum of the scaling spectra gives the topological entropy $K_0        
= \ln2$. Observe that the spectrum $g(\alpha)$ acquires also negative        
values. This does not contradict the interpretation of the scaling        
spectrum given by Bohr and Rand \cite{BR87}, which is applicable        
for $\tilde{g}(\alpha)$.        
        
Let us recall that the Hausdorff dimension $D_0$ of $\mu_1$ is the        
same as for the standard Cantor measure $\mu_{1/2}$ (or any other        
invariant measure $\mu_a$ for $a>0$) and equals $D_0 = \ln2/\ln3        
\approx 0.631$. The generalized dimensions are given by $D_q  =        
\tilde{K}_q /\ln3$ and can be fairly approximated by         
$\overline{D}_ q$ (see Sect.~IV). In Fig.~2c we compare these        
quantities with those obtained by the "box-counting" algorithm and         
observe that the difference is very small. As expected, the generalized         
dimension decreases with the R{\'e}nyi parameter $q$         
(for example $D_1\approx 0.47$ and  $D_2\approx 0.41$), which         
confirms the multifractal property of the measure $\mu_1$ (see also         
Fig.~2d). For this measure we observed that the numerical algorithm,         
providing reliable results for $q \ge 1$, definitely ceases to work for         
negative $q$.        
        
\section{Dynamical system and IFS}        
        
Let us consider a dynamical system (quantum or classical) endowed        
with an invariant measure and a partition of the phase space. We shall         
look for an IFS with the entropy equal to the entropy of the dynamical         
system with respect to the given partition. This IFS represents, in a         
sense, the backward evolution of the system \cite{S98}. Having such         
an IFS we could apply formulae for the entropy of IFS  given in         
Sect.~III and so we would obtain a new method of computing the        
dynamical entropy of the system. We illustrate this procedure on        
two examples: the R\'{e}nyi entropy of certain 1D dynamical        
systems and the coherent states (CS) entropy of certain quantum        
systems.        
        
\subsection{One-dimensional dynamical systems}        
        
Let us consider a piecewise continuously differentiable map $f :        
[0,1] \to [0,1]$. We assume that there exist subintervals $A_i$ ($i =        
1,\dots,k$) such that $[0,1] = \bigcup_{i=1}^k A_i$, $f(A_i) =        
[0,1]$, and $\left| f' \right| > 0$ in the interior of $A_i$, for each $i$.        
Let us suppose that $f$ admits an absolutely continuous invariant        
measure $\mu$ and let us denote its density by $\rho$. The partition        
$\{A_i\}_{i=1}^k$ is generating in this case, i.e., the dynamical        
entropy with respect to this partition is equal to the dynamical        
entropy of the system. With the map $f$ we can associate an IFS        
${\cal F} = \{F_i,  p_i:i=1,\dots,k\}$ given by        
        
\begin{equation}        
F_i(x) = f|_{A_i}^{-1}(x)        
\label{1d1}        
\end{equation}        
and        
        
\begin{equation}        
p_i(x) = {\rho(F_i(x)) \over  \rho(x)}  \left|F_i'(x)\right|        
\label{1d2}        
\end{equation}        
for $x \in [0,1]$ and $i=1,\dots,k$ (this is a particular case of the        
general construction from \cite{LW93,GB95}). Note that the functions         
$(F_i)_{i=1}^k$ are just continuous branches of the inverse of $f$         
(see Fig.~3).        
        
It is well known that the measure $\mu$ is also invariant for the IFS        
$\cal F$ \cite{BDEG88,IG90,GB95}. Clearly, the generalized        
entropies (given by (\ref{ren2}) and (\ref{ren3})) are in both cases        
equal, as the probabilities $P_{\mu}(i_1,\dots,i_n )$ are the same. In        
general, the most difficult stage in this construction is to show that the        
IFS $\cal F$ satisfies the assumptions which guarantee the truthfulness        
of formulae (\ref{ren5}) and (\ref{ren6}).        
        
In the present paper we analyze three examples: asymmetric {\sl "tent"        
map}, {\sl "igloo" map} (better known as the {\sl logistic map}) and        
{\sl "hut" map} given by:        
        
\vspace{1 mm} a) {\sl Tent map}: $f(y)= y/r$ for $0\le y<r$ and        
$f(y)= (1-y)/(1-r)$ for $r\le y\le1$ (where $r \in (0,1)$ is a        
parameter) with the constant invariant density $\rho =1$. Then        
$F_1(x) = rx$, $F_2(x) = (r-1)x+1$,  $p_1(x) = r$, and $p_2(x) = 1-        
r$ for $x \in [0,1]$ (Fig.~3a,b);        
        
\vspace{1 mm} b) {\sl Igloo map}: $f(y)= 4 y(1-y)$ for $y \in        
[0,1]$. In this case the invariant density has the form $\rho(y) =        
1/(\pi \sqrt{y(1-y)})$ for $y \in [0,1]$. Then  $F_1(x) = (1-\sqrt{1-        
x})/2$, $F_2(x) = (1+\sqrt{1-x})/2$, and $p_1(x) = p_2(x) = 1/2$        
for $x \in [0,1]$ (Fig.~3c,d);        
        
\vspace{1 mm} c) {\sl Hut map}: $f(y) = (-1+\sqrt{9-        
16\left| y-1/2 \right|} )/2$ for $y \in [0,1]$. The invariant density is        
given by $\rho(y)=y+1/2$ for $y \in [0,1]$. Then  $F_1(x) =        
(x^2+x)/4$, $F_2(x) = 1- ((x^2+x)/4)$, $p_1(x) = (x^2+x+2)/8$,        
and $p_2(x) = (6-x^2-x)/8$  for $x \in [0,1]$ (Fig~3e,f).        
        
\vspace{1 mm} It is easy to show that all the required assumptions        
are satisfied here and we can use formulae (\ref{ren5}) and        
(\ref{ren6}) to calculate the entropy:        
        
\vspace{1 mm} a) {\sl Tent map}: $K_q = \tilde{K}_q =\Bigl(\ln        
(r^q+(1-        
r)^q)\Bigr)/(1-q)$ for $q \neq 1$, and $K_1 = -(r \ln r+(1-r) \ln (1-        
r))$;        
        
b) {\sl Igloo map}: $K_q = \tilde{K}_q =\ln 2$ for $q \in {I\!\!R}$;        
        
 c) {\sl Hut map}: $K_q = \Bigl(\ln(4^{-q}\frac{3^{q+1}-        
1}{q+1})\Bigr)/(1-q)$ for $q \neq 1$ and $K_1 = 1/2 + 2\ln 2 -        
(9/8)\ln 3$.        
        
\vspace{1 mm} Note that, for the hut map, one can hardly obtain        
such an analytical formula for the alternative version of R\'{e}nyi        
entropy $\tilde{K}_q$.        
        
Clearly, $K_0 = \ln 2$ for each of the three maps. In the cases        
a) and c) $D_q = 1$ for each $q$. The dependence of $D_q$ on $q$        
in the case b) is presented, e.g., in \cite{S88}.        
        
A similar technique can be applied to other classes of 1D maps like,         
for example, "cookie-cutters" introduced by Bohr and Rand in         
\cite{BR87,R89}. Let us consider, e.~g., a repeller given on the unit         
interval by $f(y)=3y$ for $y\in [0,2/3]$ and $f(y)=3y-2$ for         
$y\in [2/3,1]$, for which typical (with respect to the Lebesgue measure)         
trajectories eventually leave the interval with probability one. Then the         
measure "uniformly" localized on the Cantor set is the invariant measures         
for this system. The corresponding IFS given by         
$\{F_1(x)=x/3$, $F_2(x)=(x+2)/3\}$ for $x \in [0,1]$ with constant         
probabilities $p_1 = p_2 \equiv 1/2$ is just the IFS we discussed in         
Sect.~V.        
        
\subsection{Quantum systems}        
        
In papers \cite{SZ94,ZS95} we introduced the notion of {\sl        
coherent states (CS) quantum entropy}. This quantity may be used to
 characterize chaos in quantum dynamical systems.
 Out of entire spectrum of R{\'e}nyi-like quantum 
entropies $K_q$ \cite{KSZ97}, a special meaning may be attached 
to $K_1$. Namely, the CS--entropy $K_1$ corresponds to the 
classical KS-entropy. The average of $K_1$ over the set of all
 structureless quantum systems, represented by unitary matrices 
distributed uniformly with respect to the Haar measure on $U(N)$, 
diverges with the matrix size $N$ as ln$(N)$ \cite{SZ98}. 
This result provides an argument in favor of the
 ubiquity of chaos in classical mechanics (which corresponds to 
the limit $N \to \infty$). 

The method of computing the CS--entropy based on the notion of IFS 
was proposed in \cite{KSZ97,S96} (but see also \cite{FNS91}).
 Again we have shown that one can associate with a quantum system 
and a partition of the phase space an IFS with the same entropy.        
        
Here we present an exemplary IFS obtained for the family of spin        
coherent states, the identity evolution operator, the quantum number        
$j=1/2$, and the partition of the phase (which is the two-dimensional        
sphere in this case) into two hemispheres (see        
\cite{KSZ97} for details). For this IFS we have: $X=[-1,1]$,        
$F_1(x) = (-3+2x)/(6-3x)$, $F_2(x) =  (3+2x)/(6+3x)$, $p_1(x) =        
1/2-x/4$, and $p_2(x) = 1/2+x/4$  for $x \in X$. Large contraction        
coefficient characteristic for this IFS ensures fast convergence of        
integrals performed over the measures approximating corresponding        
invariant measure. It enables us to evaluate numerically the entropy        
with an enormous precision. For example, the entropy  $K_1\approx        
0.66131433271130$, being a quantum counterpart of the classical KS-entropy, is evaluated by the deterministic algorithm        
(\ref{IFS4}) with $14$ significant digits. Such precision could be        
hardly obtained either with random iterated algorithm (\ref{IFS5})        
or with standard techniques of time series analysis        
\cite{GP83b,PS87}. The quantities characterizing the invariant        
measure of the IFS: a) R{\'e}nyi entropies: $K_q$, $\tilde{K}_q$;         
b) scaling spectra: $g(\alpha)$, $\tilde{g}(\alpha)$; c) fractal         
dimensions: $\overline{D}_ q$, $D_q$; and d) multifractal spectrum        
$f(\alpha)$ are presented in Fig.~4. The common maximum        
of the scaling spectra gives the topological entropy $K_0 = \ln 2$,        
while the same curves intersects the bisectrix at the KS-entropy        
$K_1$. Fractal dimensions $D_q$ are computed with the aid of        
the "box-counting" algorithm and compared with the quantities        
$\overline{D}_ q$ defined in Sect.~IV.        
        
\section{Concluding remarks}        
        
We have analyzed properties of IFSs with place-dependent probabilities and showed that their invariant measures often posses the {\sl multifractal property}, i.e., the fractal dimension $D_q$ depends substantially on the R{\'e}nyi parameter        
$q$.        
        
We have described a method of computing the generalized entropy        
for such IFSs by integrating the entropy function over their invariant        
measures. For numerical evaluation of the entropy one can apply the        
deterministic algorithm (useful for small number of functions) or        
random iterated algorithm (advantageous for large number of        
functions in IFS). Numerical calculations performed for generalized        
Cantor measures have shown superiority of both methods with        
respect to the standard method of computing entropy from time        
series generated by IFSs \cite{GP83b,PS87}. The entropy and the        
dimension of some IFSs of the second kind studied here display        
non-trivial scaling properties. The invariant measure for such an IFS        
may be thus neither uniform nor balanced.        
        
It is possible to attach an IFS of the second kind to certain dynamical        
systems in such a way that the generalized entropies of their        
invariant measures are equal. This idea allow us to propose a new        
method of computing entropy for dynamical systems. In this work we        
demonstrated its usefulness for some classical (R{\'e}nyi-type        
entropy of asymmetric tent map, logistic map, and "hut" map) and        
quantum (CS-measurement entropy for two hemispheres, $j=1/2$)        
systems. The method of computing entropy by integration over fractal measure has been recently applied to other dynamical systems. The tent map with a gap, related to physical problem of communication with chaos, was studied in \cite{ZB98}, while the fractal structure of an exemplary repelling system has been analyzed in \cite{LZG99}. Moreover, we used a similar method to compute the dynamical entropy of some systems with stochastic perturbations \cite{OPSZ99}. This technique may be extended for a wider class of classical and quantum dynamical systems (or even for         
Markov chains). Such results will be presented in a forthcoming publication \cite{S98}.

During last few years        
we enjoyed fruitful collaboration with late Marcin Po{\'z}niak.        
It is also a pleasure to thank Iwo Bia{\l}ynicki-Birula, {\L}ukasz        
Turski and Daniel W{\'o}jcik for inspiring discussions on integration         
over the fractal measures and for indicating the formula  (\ref{prod}).         
One of us (K.\.Z.) is thankful to Ed Ott for hospitality during his stay         
at the University of Maryland and acknowledges the Fulbright        
Fellowship.  Financial support by the Polish Committee of Scientific         
Research under grant No.~P03B~060~013 is gratefully acknowledged.

        
        
\newpage
        
\begin{figure}%
\caption{First three iterations of the uniform density on $X=[0,1]$        
by two IFSs $\{ F_1(x)=x/3, F_2(x)=(x+2)/3\}$        
attracting to the Cantor set. Figures a), b), and c) are obtained for IFS        
($a=1/2$) with constant probabilities $p_1(x)=p_2(x)=1/2$, while        
figures d), e), and f) for IFS ($a=1$) with place-dependent        
probabilities $p_1(x)=x, \ p_2(x)=1-x$.}        
\label{ff1}        
\end{figure}        
        
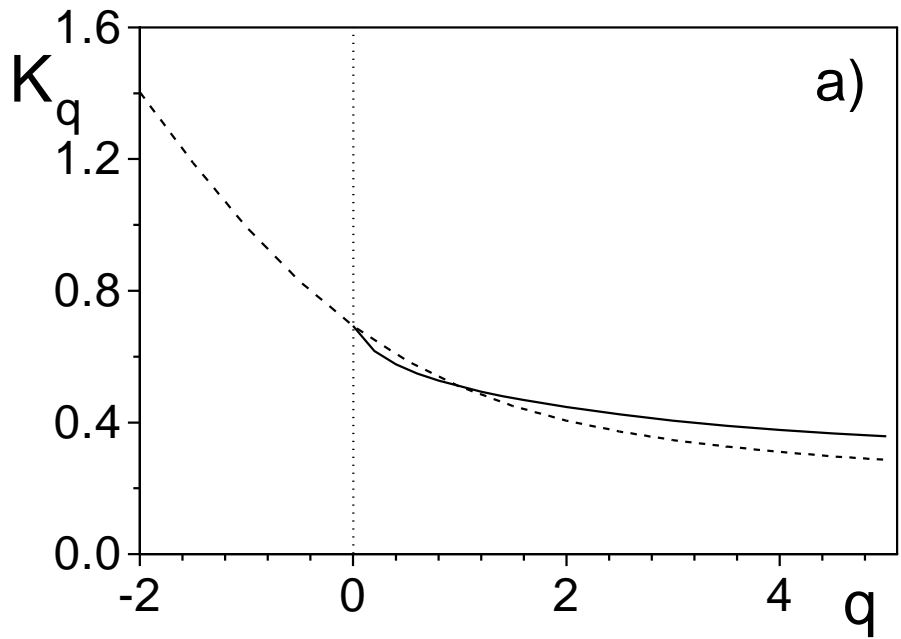
\begin{figure}%
\caption{Quantities characterizing the invariant measure for        
"Cantor" IFS of the second kind ($a=1$): a) R{\'e}nyi entropies        
$K_q$ (dashed line), $\tilde{K}_q$ (solid line); b) scaling        
spectra $g(\alpha)$ (dashed line), $\tilde{g}(\alpha)$ (solid line);        
c) fractal dimensions $\overline{D}_ q$ (solid line), $D_q$        
(stars); d) multifractal spectrum $f(\alpha)$.}        
\label{ff2}        
\end{figure}        
        
\begin{figure} %
\caption{Attaching an IFS to a 1D dynamical system:         
a) the tent map, and b) functions $F_1$ and $F_2$ of the        
corresponding IFS; c) and d) analogous pictures for the igloo        
(logistic) map; e) and f) analogous pictures for the hut map.}        
\label{ff3}        
\end{figure}        
        
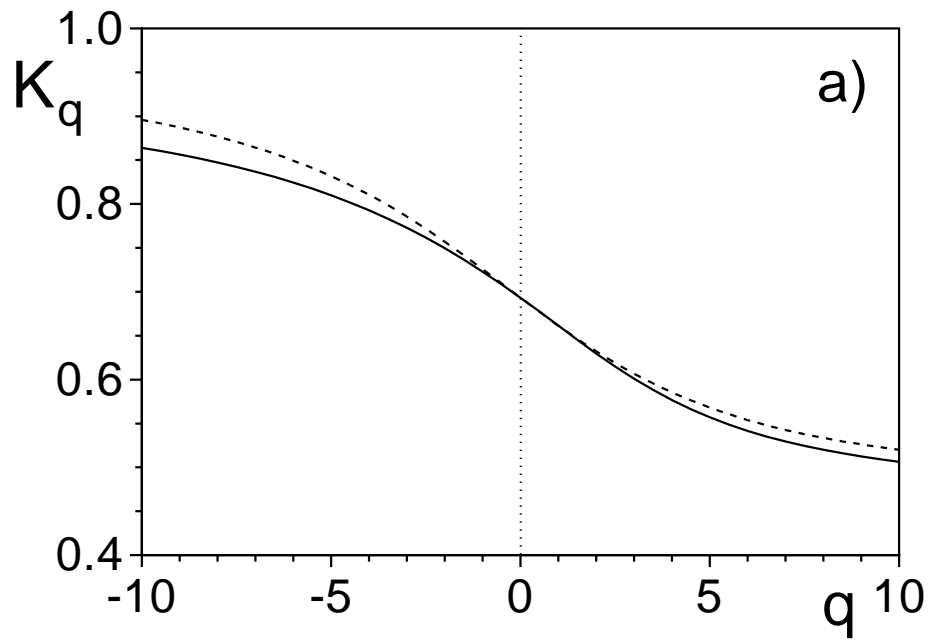
\begin{figure}        
\caption{Quantities characterizing the invariant measure of the IFS        
related to the quantum system: a) R{\'e}nyi entropies        
$K_q$ (dashed line), $\tilde{K}_q$ (solid line); b) scaling        
spectra $g(\alpha)$ (dashed line), $\tilde{g}(\alpha)$ (solid line);        
c) fractal dimensions $\overline{D}_ q $ (solid line), $D_q$        
(stars); d) multifractal spectrum $f(\alpha)$.}        
\label{ff4}        
\end{figure}        
        
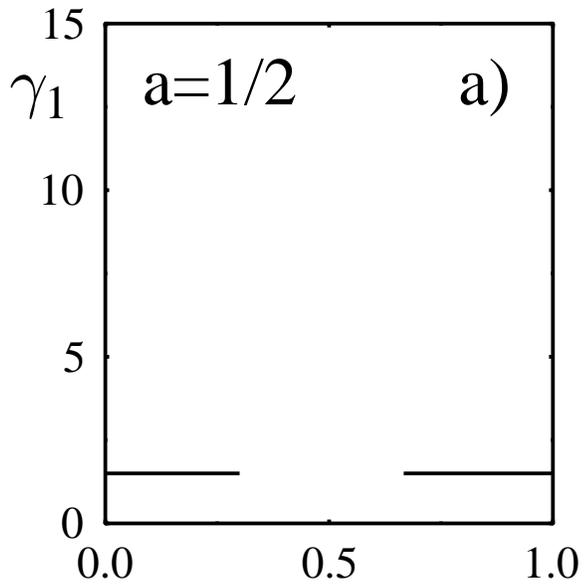
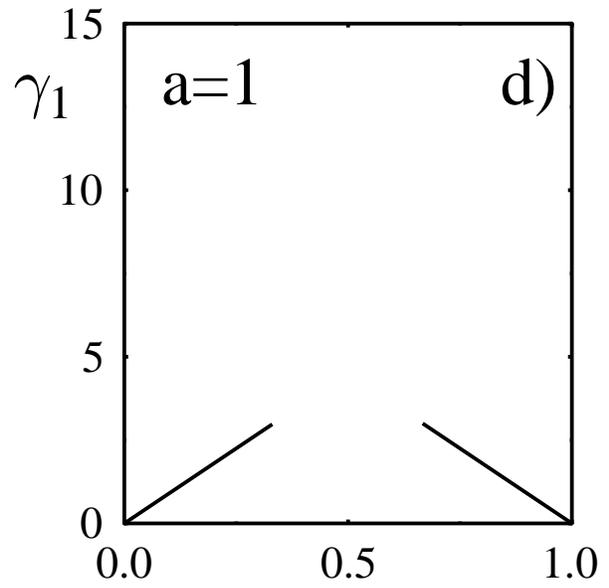
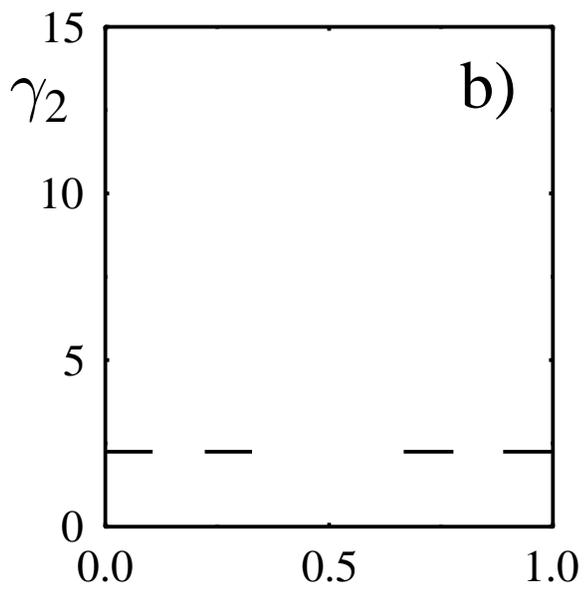
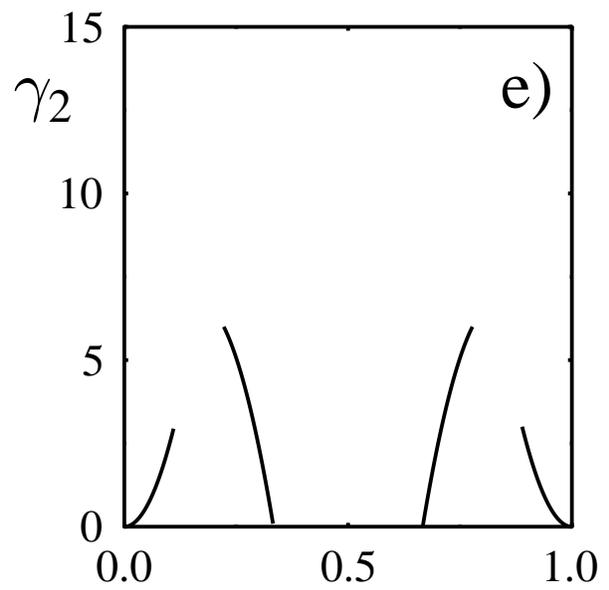
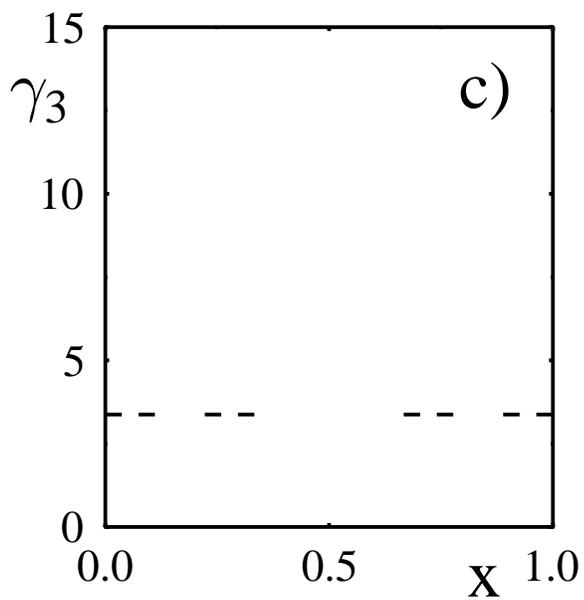
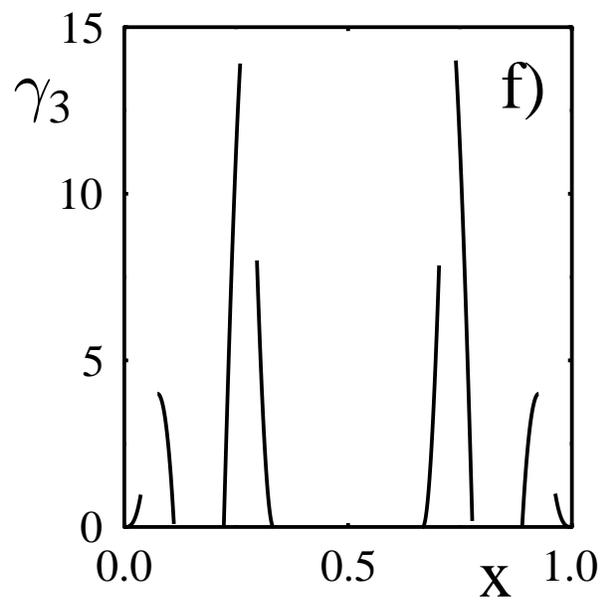
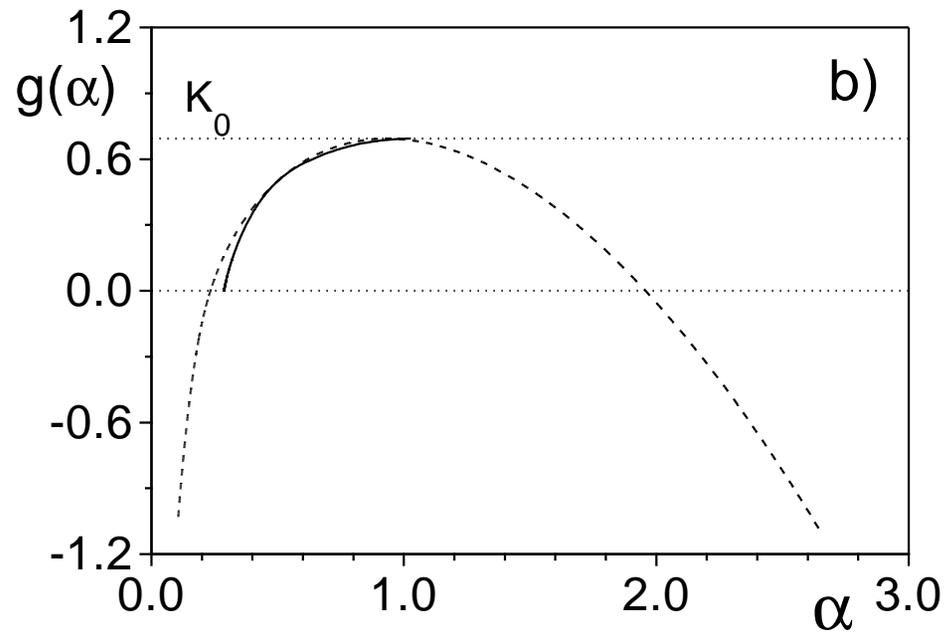
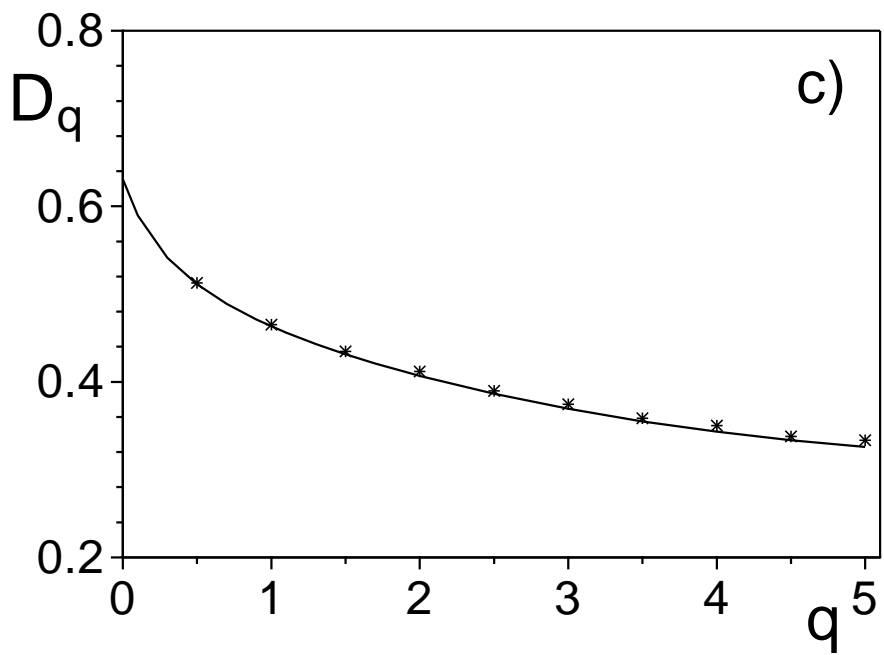
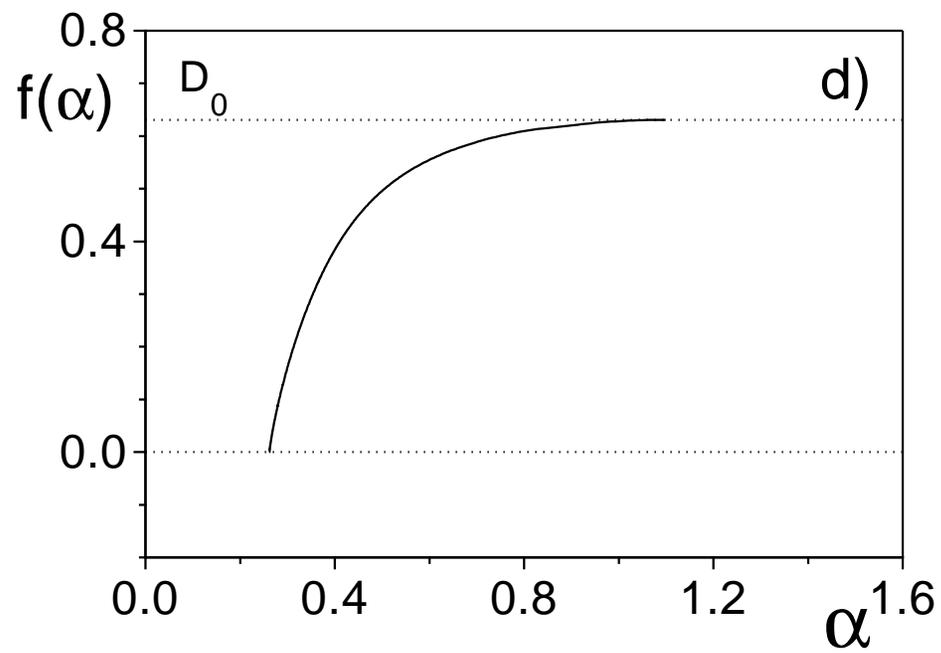
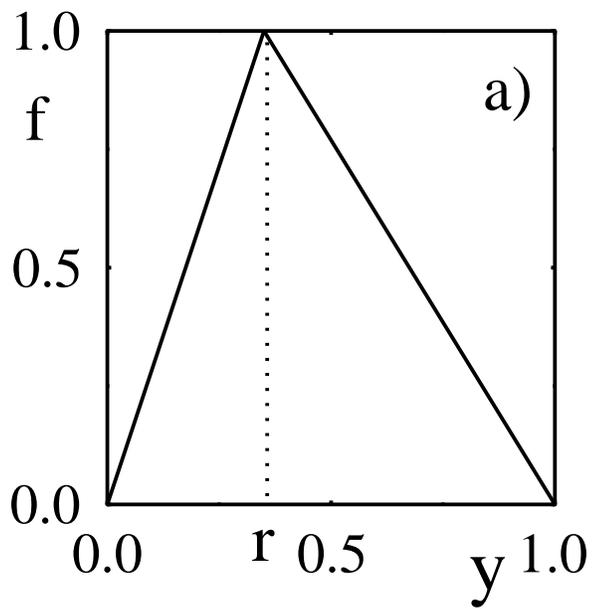
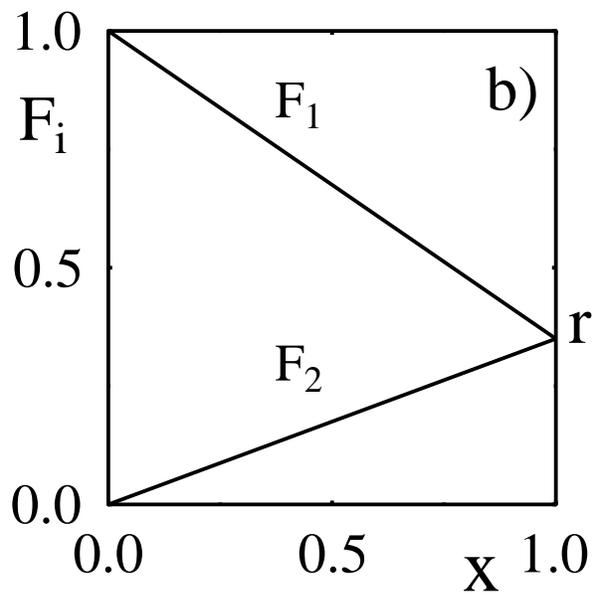
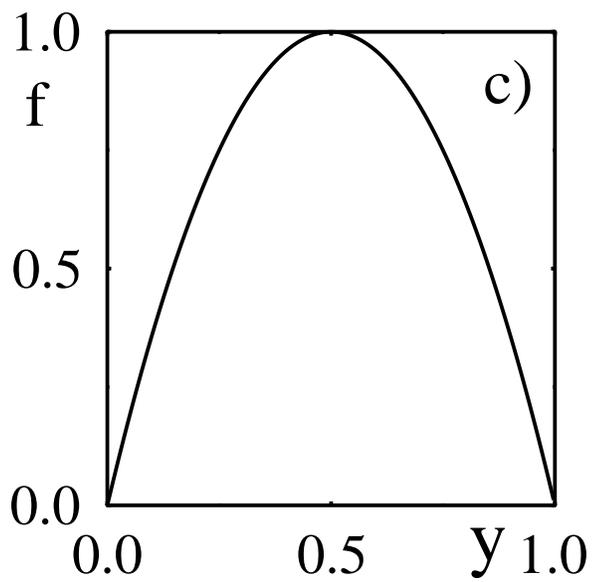
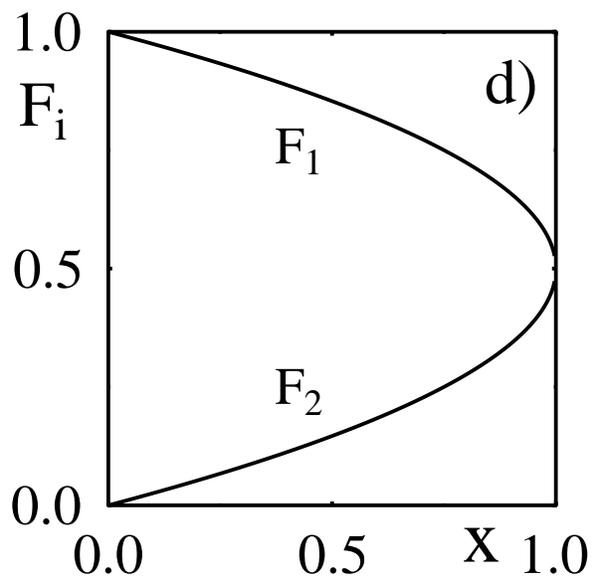
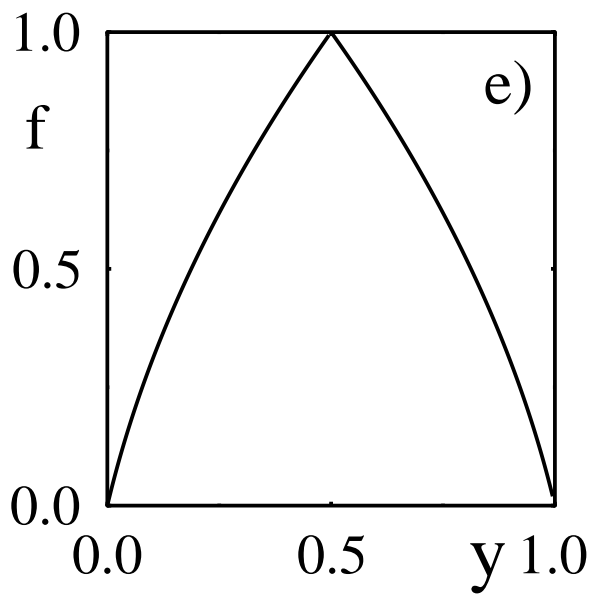
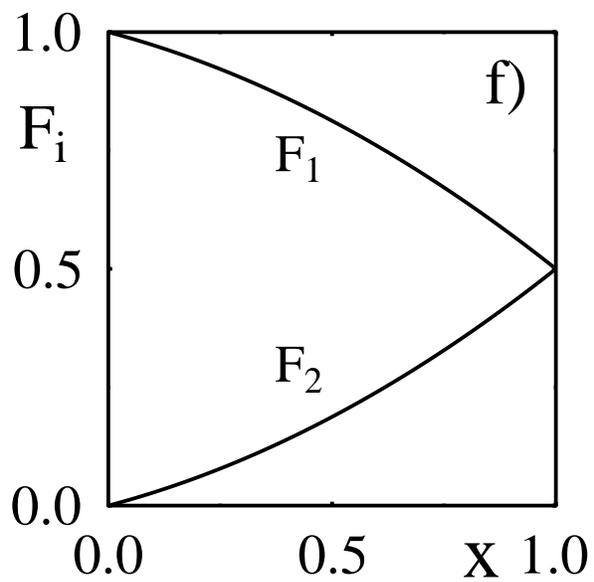
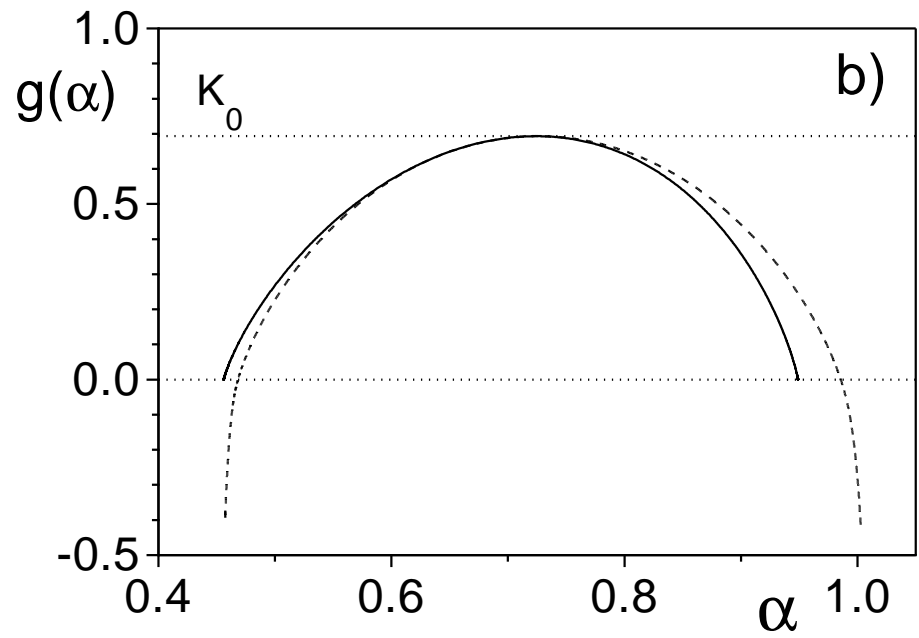
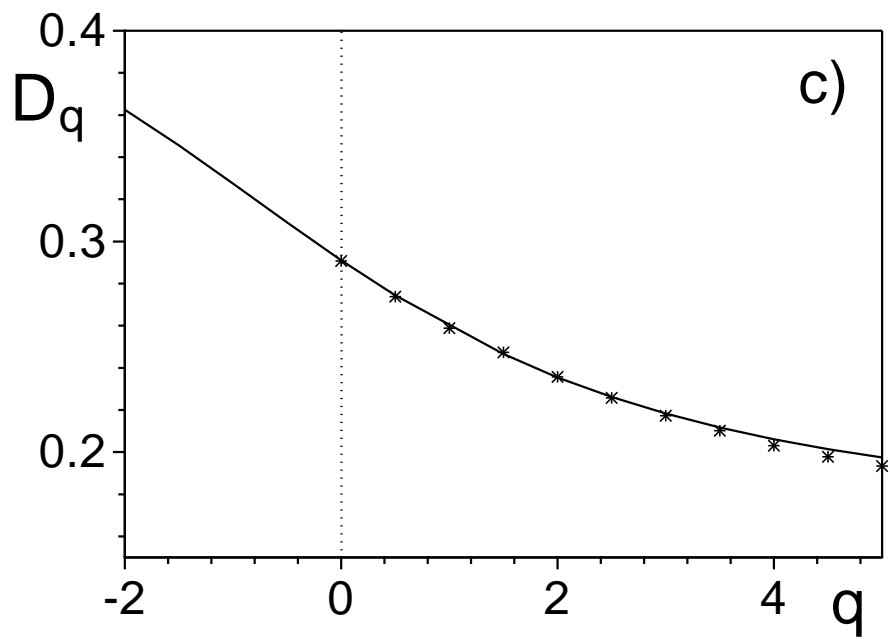
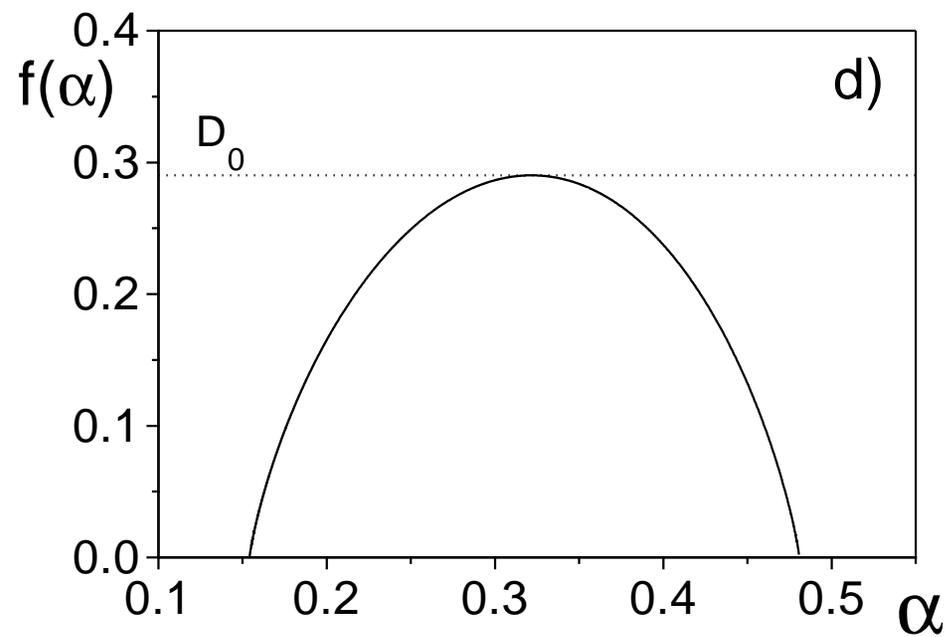
\end{document}